\documentclass[iop]{emulateapj}
\usepackage{graphicx} \usepackage{natbib} 
\shorttitle{Evidence for a $\sim 300$ Mpc Scale Local Under-density}

\begin{document}

\title{Evidence for a $\sim 300$ Megaparsec Scale Under-density in the Local Galaxy Distribution}
\author{R. C. Keenan\altaffilmark{1}, A. J. Barger\altaffilmark{2,3,4},
L. L.Cowie\altaffilmark{4}}

\altaffiltext{1}{Academia Sinica Institute of Astronomy and Astrophysics, P.O. Box 23-141, Taipei 10617, Taiwan} 
\altaffiltext{2}{Department of Astronomy, University of Wisconsin-Madison, 475 North Charter Street, Madison, WI 53706, USA}
\altaffiltext{3}{Department of Physics and Astronomy, University of Hawaii, 2505 Correa Road, Honolulu, HI 96822, USA}
\altaffiltext{4}{Institute for Astronomy, University of Hawaii, 2680 Woodlawn Drive, Honolulu, HI 96822, USA}

\begin{abstract}
Galaxy counts and recent measurements of the luminosity density in the near-infrared have indicated the possibility that the local universe may be under-dense on scales of several hundred megaparsecs.  The presence of a large-scale under-density in the local universe could introduce significant biases into the interpretation of cosmological observables, and, in particular, into the inferred effects of dark energy on the expansion rate.  Here we measure the $K-$band luminosity density as a function of redshift  to test for such a local under-density.  For our primary sample in this study, we select galaxies from the UKIDSS Large Area Survey and use spectroscopy from the SDSS, 2DFGRS, GAMA, and other redshift surveys to generate a $K-$selected catalog of $\sim 35,000$ galaxies that is $\sim 95\%$ spectroscopically complete at $K_{\rm{AB}}< 16.3$ ($K_{\rm{AB}}< 17$ in the GAMA fields).  To complement this sample at low redshifts, we also analyze a $K-$selected sample from the 2M++ catalog, which combines 2MASS photometry with redshifts from the 2MASS redshift survey, the 6DFGRS, and the SDSS.  The combination of these samples allows for a detailed measurement of the $K-$band luminosity density as a function of distance over the redshift range $0.01 < z < 0.2$ (radial distances $D\sim 50-800$~$h_{70}^{-1}$~Mpc).  We find that the overall shape of the $z=0$ rest-frame  $K-$band luminosity function ($M^* = -22.15\pm 0.04$ and $\alpha=-1.02\pm 0.03$) appears to be relatively constant as a function of environment and distance from us.   We find a local ($z < 0.07, D < 300$~$h_{70}^{-1}$~Mpc) luminosity density that is in good agreement with previous studies.  Beyond $z\sim 0.07$, we detect a rising luminosity density that reaches a value of roughly $\sim 1.5$ times higher than that measured locally at $z>0.1$.   This suggests that the stellar mass density as a function of distance follows a similar trend.    Assuming that luminous matter traces the underlying dark matter distribution, this implies that the local mass density of the universe may be lower than the global mass density on a scale and amplitude sufficient to introduce significant biases into the determination of basic cosmological observables.  An under-density of roughly this scale and amplitude could resolve the apparent tension between direct measurements of the Hubble constant and those inferred by Planck.    
\end{abstract} \keywords{cosmology: observations --- cosmology: large-scale structure of universe --- galaxies: fundamental parameters --- galaxies: luminosity function}
\maketitle

\section{Introduction}
\label{intro}

The universe is generally assumed to be isotropic and homogeneous on very large scales.  This allows for the development of cosmological models and observations to constrain those models, provided that the observations are made over a sufficiently large volume to average over so called ``cosmic variance", or systematic measurement biases due to large-scale structure.  Observations of the cosmic microwave background (CMB) indicate the universe was very homogeneous at $z\sim 1100$, and measurements of the kinematic Sunyaev-Zeldovich (kSZ) effect appear to indicate that the present day universe is homogeneous on scales greater than $\sim 1$~Gpc \citep{Garc08b, Zhan11,Ade13ksz}.  However, the large-scale homogeneity of the universe on smaller scales has not been measured directly.  

Observed large-scale structures, such as the $>400~$Mpc Sloan Great Wall \citep{Gott05}, demonstrate the existence of inhomogeneity on very large scales.  In addition, it has been shown that under-densities on similarly large scales may explain the ``cold spots'' in the CMB \citep{Inou06}.  While such structures had previously been thought to be departures from the expectation of Lambda-Cold-Dark-Matter ($\Lambda$CDM) cosmology \citep{Shet11}, it has now been found that such large-scale structures are not only to be expected in the current concordance cosmology, but also that still larger structures are likely to be discovered as survey volumes increase \citep{Park12}.  

Recent cosmological modeling efforts have demonstrated that large-scale structure in the local universe may introduce significant systematic errors into the measurement of cosmological observables and hence the interpretation of these observables in the context of a given cosmological model  (for recent reviews see \citealt{Bole11b, Clar12a} ).  In particular, so called ``void models" place the observer inside a large local under-density to provide for the apparent acceleration of the expansion of the universe \citep{Moff92,Moff95,Cele00,Tomi00,Tomi01a,Tomi01b,Iguc02,Alne06, Chun06, Enqv07, Yoo08,Garc08a,Alex09,Garc09,Febr10,Cele10, Bisw10,Marr10,Clar10,Bole11a,Nish12,Valk12a,Roma11,Roma12}. 

The basic idea underlying these void models is that if an observer lives near the center of a large under-density, then that observer will witness a local expansion of the universe that is faster than the global expansion. This would result in a locally measured Hubble constant that is larger than the global expansion rate and appear observationally as an accelerating expansion.

In their current form, void models without a cosmological constant do not appear to be viable alternatives to dark-energy-dominated universes, as they have problems in simultaneously fitting all cosmological observables  \citep{Garc08b, Zibi08, Moss11, Zhan11, Ries11,Zuma12}. However, the exploration of these types of cosmological models---and other models which explore the effects of large-scale inhomogeneity---has highlighted the need for a more thorough understanding of extremely large-scale structure in the local universe \citep{Marr11, Marr12,Marr13,Bull12,Mish12, Mish13, Valk12b, Valk13,Roma10,Roma13}. 

In particular, ``minimal void" scenarios (e.g., \citealt{Alex09, Bole11a}) have shown that very simple models placing the observer near the center of an under-density that is $\sim 300$ Mpc in radius and roughly half the density of its surroundings are sufficient to explain the acceleration observed via type Ia supernovae. While these models are simplistic, they make clear that an observer's location with respect to large-scale structures could have profound implications for that observer's measurement of cosmological observables.

In this study, we wish to make an estimate of the mass density of the nearby universe as a function of distance from us to test for large-scale inhomogeneity, and, in particular, for a large local under-density.  While we cannot directly probe the underlying dark matter distribution, we can make a robust estimate of the near-infrared (NIR) luminosity density, which is a good tracer of the overall stellar mass density (e.g., \citealt{Dejo96, Bell01, Bell03, Kirb08}).   

The stellar masses of galaxies, on average, have been shown to be correlated with the mass of their host dark matter halos (e.g., \citealt{WangL12}).  Furthermore, simulations have shown that on much larger ($\sim 100$~Mpc) scales at low redshift, the spatial distribution of baryons should be an unbiased tracer of the underlying dark matter distribution \citep{Angu13}.  

Thus, in terms of the linear bias parameter, here we will assume $b=1$ in the relation $\delta_g = b\delta$, where $\delta_g$ is the relative density contrast given by the distribution of galaxies, i.e., $(\rho_{stars,z}-\overline{\rho}_{stars}) / \overline{\rho}_{stars} $, and $\delta$ is the total mass density contrast, $(\rho_z - \overline{\rho}) / \overline{\rho}$.    At optical wavelengths, the linear bias parameter has been observed to approach unity at low redshifts (e.g., \citealt{Mari05}), and in the NIR, \citet{Mall05} have measured a value in the $K-$band of $b_K(z\approx 0) = 1.1 \pm 0.2$, suggesting the NIR bias parameter follows a similar trend.   

   Therefore, a measurement of the NIR luminosity density can provide an estimate of the underlying mass density of the universe.  Likewise, a measured change in NIR luminosity density as a function of redshift could signal a corresponding change in the underlying total mass density.   Thus, in the NIR at low redshifts, where dust extinction is minimal and $K-$corrections are small and nearly independent of galaxy type, statistical studies of galaxies provide a means of probing local large-scale structure.    

\begin{figure*}
\begin{center}
\includegraphics[width=180mm]{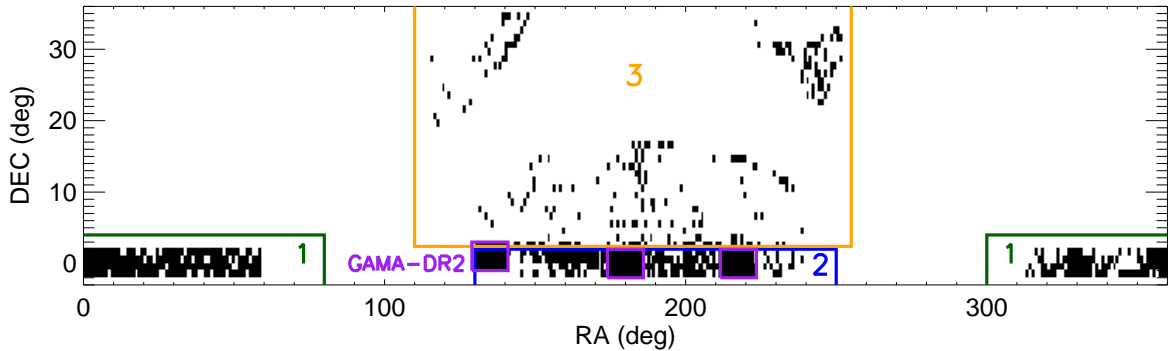}
\caption{\label{cover} The region on the sky where the UKIDSS-LAS DR8 and the SDSS DR9 overlap and the spectroscopic completeness is $\sim 95\%$ at $K_{\rm{AB}}<16.3$ ($\sim 585.4$ deg$^2$ total). The black rectangles represent $1^\circ\times 1^\circ$ areas of high completeness within the overlap region.  The three purple boxes represent the areas covered by the GAMA survey second data release ($144$ deg$^2$ complete to $K_{\rm{AB}}=17$).  The boxes drawn in green, blue, and orange denote the three subregions in this sample that are subsequently dealt with separately to address possible biases and cosmic variance in this study.   }
\end{center}
\end{figure*} 

Several studies of NIR galaxy counts have found that the local space density of galaxies appears to be low by $\sim 25-50\%$ compared to the density at distances of $\gtrsim 300$ Mpc \citep{Huan97, Frit03,Frit05a, Frit06,Buss04,  Keen10a,Whit13}. A similar result has also been noted in studies of galaxy counts and galaxy space densities in optically selected samples \citep{Madd90,Zucc97}.  If the space density of galaxies is rising as a function of redshift, then a corresponding rise in luminosity density should also be present.   

In \citet{Keen12}, we probed the NIR luminosity function (LF) just beyond the local volume at redshifts of $z\sim 0.2$. We found that the NIR luminosity density at $z\sim 0.2$ appears to be $\sim 30\%$ higher than that measured at $z\sim 0.05$.  This measured excess could be considered a conservative underestimate, because we avoided known over-densities, such as galaxy clusters, in our study.  We note that our result cannot be considered conclusive, given possible systematics due to cosmic variance in our measurement.  However, taken in the context of other measurements of the NIR luminosity density from the literature, our result is consistent with other studies that found a higher luminosity density at $z>0.1$ than that found locally.  Furthermore, we showed that all measurements of the luminosity density could be considered roughly consistent with the radial density profile from \citet{Bole11a}, which they claim could mimic the apparent acceleration of the expansion of the universe.

In the past, wide-area NIR surveys have generally not gone deep enough to probe beyond the very local universe, and deep surveys have been carried out over relatively small solid angles, such that good counting statistics cannot be achieved at lower redshifts.  The result is that, in a study such as that performed in \citet{Keen12}, we are comparing measurements made with quite different photometry and methodology, and the results may suffer from unknown biases.  

Recently, however, the completion of the relatively deep and wide UKIRT Infrared Deep Sky Large Area Survey (UKIDSS-LAS, \citealt{Lawr07}), combined with spectroscopy  from the Sloan Digital Sky Survey (SDSS, \citealt{York00}), the Two-degree Field Galaxy Redshift Survey (2dFGRS, \citealt{Coll01}), the Galaxy And Mass Assembly Survey (GAMA, \citealt{Driv09,Driv11}), and other spectroscopic data from the archives, provides for a NIR-selected spectroscopic sample of galaxies that is both wide enough on the sky and deep enough photometrically to sample a relatively broad range in redshifts in the nearby universe with good statistics.  This sample allows us to study the $K-$band galaxy LF over a large volume in the redshift range $0.01 < z < 0.2$.  With these data we are able, for the first time, to measure the $K-$band galaxy LF as a function of redshift using consistent photometry and methodology.  

To confirm our results at the low-redshift end of the UKIDSS sample, we analyze the highly spectroscopically complete all-sky sample of $K-$selected galaxies (the 2M++ catalog) compiled by \citet{Lava11}.   The 2M++ catalog combines photometry from the Two Micron All Sky Survey Extended Source Catalog (2MASS-XSC, \citealt{Skru06}) with redshifts from the Two Micron Redshift Survey (2MRS, \citealt{Huch05, Erdo06}), the Six Degree Field Galaxy Redshift Survey (6DFGRS, \citealt{Jone09}), and the SDSS.  

Thus, in this study we are leveraging all available data from the largest photometric and spectroscopic surveys to perform the most rigorous study to date of the $K-$band luminosity density in the local universe. 

The structure of this paper is the following.   We discuss our sample selection in Section~\ref{catgen}.  We estimate the $K-$band luminosity density as a function of distance in Section~\ref{klf}.  We discuss possible biases in our measurements in Section~\ref{biases}.  We summarize our results in Section~\ref{summary}.  Unless otherwise noted, all magnitudes listed here are in the AB magnitude system ($m_{\rm{AB}} = 23.9-2.5~\rm{log}_{10}~$$ f_\nu$ with $f_\nu$ in units of $\mu$Jy).  We assume $\Omega_M = 0.3, \Omega_{\Lambda} = 0.7,~$and$~H_0 = 70$~km~s$^{-1}$~Mpc$^{-1}$ (i.e., $h_{70}=1$) in our calculations.

\section{Catalog Generation}
\label{catgen}

To create a stellar mass selected sample, we have retrieved data from the WFCam Science Archive (WSA) for all galaxies with NIR photometry in the UKIDSS-LAS DR8 in the $K-$band ($K_{\rm{Vega}}<14.4$, which is equivalent to $K_{\rm{AB}}<16.3$), where overlap with the SDSS, the 2DFGRS, GAMA, and other surveys provides for highly complete spectroscopy of bright galaxies ($\sim 95\%$ complete at $K_{\rm{AB}}<16.3$, or $K_{\rm{AB}}<17$ in the GAMA fields).  This region of sky covers $585.4$~deg$^2$ (see Figure~\ref{cover}), and the selection contains $35,342$ galaxies at a median redshift of $z \sim 0.1$.  The UKIDSS-LAS $K-$band depth is $K_{\rm{AB}}\approx 20$, so this sample features very high signal to noise $K-$band photometry.

We then used the SDSS-III Sky Server CasJobs\footnote{skyserver.sdss3.org/CasJobs} interface to retrieve optical photometry and redshifts from the SDSS DR9 over the areas covered in all bands ($Y, J, H,$~and~$K$) in the UKIDSS-LAS. We then cross correlated the positions of the $K-$band selected objects with those from the SDSS using a search radius of $2\arcsec$ (currently the WSA provides its own cross-matched catalog to the SDSS-DR9, but this was unavailable at the time of these analyses).   

A small fraction ($<1\%$) of objects  identified in the UKIDSS catalogs did not have a  counterpart within $<2\arcsec$ in the SDSS.  We found that these objects generally appeared to be spurious $K-$band detections (checking by eye in the imaging data from UKIDSS), so we excluded these objects from the final catalog.  Thus, all of the UKIDSS $K-$band selected objects in our catalog have counterparts classified as primary target galaxies in the SDSS.  We find that $95\%$ of UKIDSS objects matched to SDSS counterparts have angular separations of $<0\farcs5$, and the average separation between UKIDSS objects and their SDSS counterparts is $0\farcs2$. 

We also downloaded the 2dFGRS redshift catalogs from the VizieR online service\footnote{http://vizier.cfa.harvard.edu/viz-bin/VizieR-3}.  We matched 2dF objects to their UKIDSS counterparts using  a search radius of $2\arcsec$.   Cross-correlation with the 2dFGRS significantly improved overall completeness in the 2dF equatorial region from $10^h < {\rm RA} < 15^h$ (subregion 2 denoted in Figure~\ref{cover}).  We also cross-correlated with published redshifts in the NASA-Sloan Atlas\footnote{www.nsatlas.org} and the NASA/IPAC Extragalactic Database\footnote{ned.ipac.caltech.edu} to augment the completeness of the sample.  

The GAMA collaboration recently published their second data release (DR2, J. Liske et al. 2013 in preparation) including redshifts and independent $K-$band photometry \citep{Hill11} for three target fields covering $144$ deg$^2$ within the UKIDSS-LAS area.  The GAMA $K-$band photometry is derived from UKIDSS-LAS $K-$band imaging.  We use the GAMA-DR2 redshifts to further expand the $K-$selected sample.  Within the GAMA fields themselves, the deeper spectroscopy ($\sim 98\%$ complete at $K_{\rm{AB}}<17$) allows us to push further down the faint end of the LF and resolve significantly more of the total light from galaxies at $z>0.1$.     

\begin{figure}
\begin{center}
\includegraphics[width=80mm]{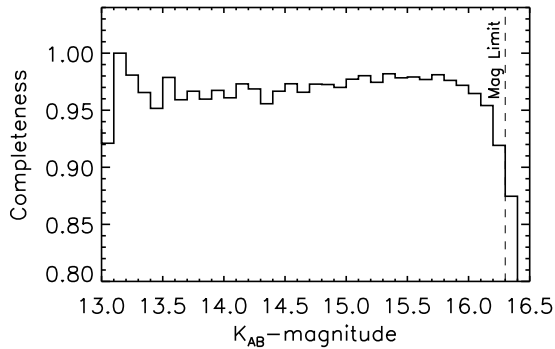}
\caption{\label{complete} Completeness as a function of $K-$band apparent magnitude for the UKIDSS sample.  The average completeness of the full sample of $35,342$ galaxies with $K_{\rm{AB}} < 16.3$ is $\sim 95\%$.}
\end{center}
\end{figure}

\begin{figure}
\begin{center}
\includegraphics[width=80mm]{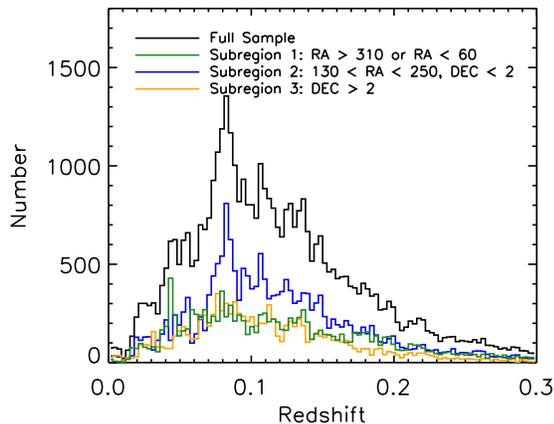}
\caption{\label{zhist} The redshift histogram for the spectroscopic sample of $K_{\rm{AB}}<16.3$ galaxies in this study.  Subregion 2 overlaps with the Sloan Great Wall identified by \citet{Gott05}, and the peak near $z \sim 0.08$ in the redshift histogram of this subregion is primarily due to this structure.}
\end{center}
\end{figure}

We restricted our final catalog to regions on the sky that are $>90\%$ spectroscopically complete at $K_{\rm{AB}} =16.3$ in order to minimize possible biases associated with the fact that our sample is selected in the $K-$band, while the targets for the surveys providing the redshifts for this work were primarily optically selected.  This resulted in a catalog of $35,342$ galaxies over the $585.4$~deg$^2$ region shown in Figure~\ref{cover}, after excluding stars from the catalog, as we describe in the following subsection.  We show the completeness of the sample as a function of apparent magnitude in Figure~\ref{complete}, and we show the redshift distribution in Figure~\ref{zhist}.

\subsection{Star-Galaxy Separation}
\label{sgsep}

\begin{figure}
\begin{center}
\includegraphics[width=80mm]{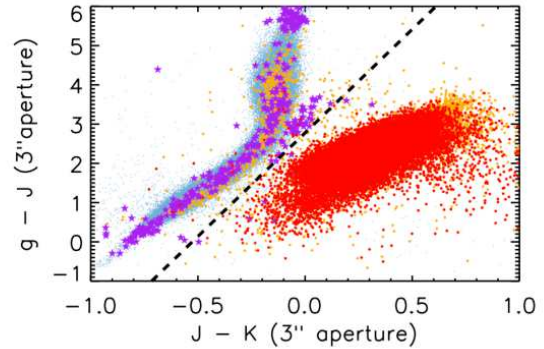}
\caption{\label{fibergjk} A $3\arcsec$ aperture magnitude color-color separation for all objects at $K_{\rm{AB}}<16.3$ in the raw sample drawn from the WSA and SDSS Skyserver.  Orange points show objects classified by morphology as extended in both UKIDSS and the SDSS (UKIDSS $mergedclass=1$ and SDSS $type=6$).  Light blue points show objects classified as point sources in both UKIDSS and the SDSS (UKIDSS $mergedclass=-1$ and SDSS $type=3$).  Red points show spectroscopically identified galaxies.  Purple stars show spectroscopically identified stars.  The dashed line shows the best color separation boundary between stars and galaxies: $g-J =  5.28\times [J-K] + 2.78$.}
\end{center}
\end{figure}

We investigated the $g-J$ vs. $J-K$ color-color diagram for objects in our catalog to determine the reliability of the star-galaxy classifiers offered in the WSA and SDSS archives (UKIDSS ``mergedclass" and SDSS ``type").   We show a $gJK$ color-color diagram for $3\arcsec$ aperture magnitude colors in Figure~\ref{fibergjk}.  Orange points show objects classified by morphology as extended in both UKIDSS and the SDSS (UKIDSS $mergedclass=1$ and SDSS $type=6$).  Light blue points show objects classified as point sources in both UKIDSS and the SDSS (UKIDSS $mergedclass=-1$ and SDSS $type=3$).  Red points show spectroscopically identified galaxies.  Purple stars show spectroscopically identified stars.  The dashed line shows the best color separation boundary between stars and galaxies: $g-J =  5.28\times [J-K] + 2.78$.   Approximately $2\%$ of extended sources lie above the separation boundary, and $\sim 0.5\%$ of point sources lie below.

Next, we looked at the images in the SDSS Sky Server database of the point sources that lie below the separation boundary  and the extended sources that lie above.  For the point sources, it was clear from the SDSS imaging that they are point-like in morphology.  We expect that most of these objects are stars with rare colors or bad photometry in one band (with the exception of a few quasars, which are not of interest for this study).  

The vast majority of the  extended sources that lie above the separation boundary lie along the stellar main sequence for this color separation.  Upon investigating the imaging for these sources, we found that, while there are definitely a few galaxies among them, the vast majority ($\sim 99\%$) look like point sources that are either smeared out a bit or blended with another object.  Thus, we conclude that most of the objects that are classified as extended but lie above the boundary are, in fact, stars.  

We note that 15 spectroscopically confirmed galaxies lie above the separation boundary.  Upon closer inspection of SDSS imaging for these sources, we found that roughly a third appeared to be point-sources on the stellar main sequence.  Another third appeared to be galaxies blended with a star, and the final third appeared to be galaxies by morphology, but with stellar colors.   There are only a handful of spectroscopically confirmed stars that lie below the separation boundary, and all appear to be point sources.  

Based on these analyses, we conclude that by excluding all sources above the separation boundary described in Figure~\ref{fibergjk}, as well as all sources identified by morphology as point sources in the SDSS and WSA (regardless of color), we can achieve a star-galaxy separation that is robust at better than the $1\%$ level.  

\subsection{Petrosian Aperture Clipping in UKIDSS Photometry}
\label{clipping}

The sky subtraction algorithm in the pipeline for UKIDSS photometry is such that there exists an upper limit of $6\arcsec$ on the Petrosian aperture radius (corresponding to a circular aperture with a radius of $12\arcsec$).  This causes the total flux from galaxies that subtend large solid angles to be underestimated.  This implies that if we use Petrosian magnitudes, then we will underestimate the luminosities of some galaxies, while we will lose other galaxies from the sample completely, because the aperture clipping pushes them to a fainter apparent magnitude than the selection limit of the sample.  

\begin{figure}
\begin{center}
\includegraphics[width=80mm]{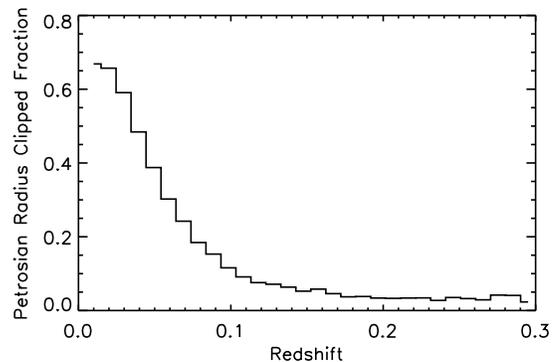}
\caption{\label{clippedvz} The fraction of galaxies with their Petrosian aperture clipped at $6\arcsec$ as a function of redshift in the UKIDSS-LAS.  We correct for this effect by extrapolating the surface brightness profile for each galaxy derived from a range of circular aperture magnitudes, as described in Section~\ref{clipping}.}
\end{center}
\end{figure}

We retrieved the $K-$band Petrosian aperture radii for our sample from the WSA and found that roughly $10\%$ of the galaxies had their Petrosian apertures clipped at $6\arcsec$.  We show the fraction of galaxies for which the Petrosian aperture was clipped at $6\arcsec$ as a function of redshift in Figure~\ref{clippedvz}.  Clearly, the underestimation of total flux is a much stronger effect at low redshift ($z<0.1$).  

Instead of omitting galaxies for which the Petrosian apertures were clipped (e.g., \citealt{Smit09}), we have devised a method to compensate for the effects of the underestimation of flux.  Short of redoing all the photometry, there is no way to know exactly what fraction of light was lost due to the aperture clipping.  However, for each galaxy the WSA provides photometry for circular apertures ranging in size from $1\arcsec$ to $12\arcsec$ in radius.  Using these measurements, we estimated a surface brightness profile for each galaxy affected by Petrosian aperture clipping by fitting a S\'{e}rsic profile to the aperture photometry.  We then extrapolated this profile out to twice the SDSS $z-$band Petrosian radius provided for each galaxy to estimate the flux lost due to aperture clipping.  In the NASA-Sloan Atlas (NSA), new and improved photometry is provided for SDSS galaxies at $z< 0.055$.  Thus, at $z<0.055$, we extrapolate to twice the new $z-$band Petrosian radius provided by the NSA.  

While this method is rather crude, it allows for a means of estimating the light lost due to aperture clipping.  We found that the median clipped aperture correction was $\sim 0.04$ magnitudes.  We applied this method to all galaxies with clipped apertures down to 2 magnitudes below our selection limit  for this study ($K_{\rm{AB}}<16.3$).  We found that this correction only increased the total number of galaxies in the sample by $\sim 0.25\%$ (via the movement of galaxies from fainter to brighter than the magnitude selection limit after the aperture correction).   We conducted all of the analyses that follow both with and without this correction applied and found essentially no change in our results.    

As an additional check, we compare with the new UKIDSS-LAS $K-$band photometry provided by the GAMA collaboration \citep{Hill11} for objects in their survey fields.  The GAMA Petrosian apertures used for $K-$band photometry are matched to the apertures derived from $R-$band source extraction.  While this new photometry is not strictly equivalent to Petrosian aperture photometry defined from $K-$band source extraction, the GAMA Petrosian radii are not arbitrarily clipped, so we have a means of comparing clipped and unclipped photometry for a subsample of the UKIDSS catalog.

We find that for objects not affected by aperture clipping in the UKIDSS sample, there is an rms scatter of $\sim 0.1$ mags between GAMA and UKIDSS Petrosian magnitudes and that GAMA magnitudes are systematically brighter by $\sim 0.03$ mags.  For (uncorrected) clipped objects, we find the same scatter, but that the GAMA magnitudes are $\sim 0.07$ mags brighter.  The aperture clipping correction described above results in the systematic magnitude offset for clipped objects being reduced to $\sim 0.03$ mags, in agreement with that for unclipped objects, so we take this as further evidence that the clipping correction is appropriate.   

\subsection{Area Estimation}

The exact estimate of the area is not critical to the results presented below, as it only serves to set the overall normalization.  However, given that we wish to investigate the surface density and volume density of galaxies, we need to have an estimate the area on the sky (and hence volume) of our survey.  The UKIDSS-LAS footprint features gaps and holes on a range of scales that make the area estimation a nontrivial task.  

To deal with this issue, we employ a ``counts-in-cells" method of determining area coverage.   In our initial catalog of galaxies and stars, the median angular separation between nearest neighbors is $\approx 150\arcsec$.    Thus, we divide the entire area surveyed into square cells $ 150\arcsec \times 150\arcsec$ in size.  From the center of each cell we determine whether there is a catalogued object within a radius of twice the characteristic separation between objects ($300\arcsec$).  If an object is found then the cell is counted as surveyed area.

       We then simply counted up the number of surveyed cells to compute the total area surveyed.  Using this method, we computed a total area for the overlap region between SDSS and UKIDSS of $1952.1$ deg$^2$ on the sky.  Next we take the source counts in the areas of high completeness ($>90\%$ at $K_{\rm{AB}}<16.3$) divided by the total source counts to estimate the area to be considered in this study ($585.4~$deg$^2$) Varying our cell size and search radius by a factor of 2 changed our area estimate by only $\sim 2\%$.  Thus, we estimate  our error on the area calculation is $\sim 2\%$.

\subsection{Galaxy Counts as a Catalog Check}
\label{catcheck}
\begin{figure}
\begin{center}
\includegraphics[width=90mm]{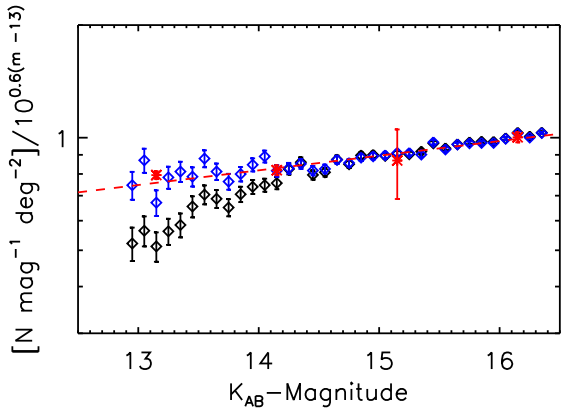}
\caption{\label{counts} $K-$band galaxy counts as a function of apparent magnitude for this study compared with average $K_s-$band galaxy counts from \citet{Keen10b}.  The counts have been divided through by a normalized Euclidean model of slope $\alpha = 0.6$ to expand the ordinate (assuming the bright galaxy counts take the form $N(m) = A\times 10^{\alpha m}$, where $A$ is a constant).  Counts from \citet{Keen10b} are shown as red asterisks with a fit to these counts shown as a red dashed line.  Raw counts from this study are shown as black diamonds.  Counts from this study after the aperture clipping correction was applied are shown as blue diamonds.  Note: we shift the \citet{Keen10b} $K_s-$band counts by $+0.15$ in magnitude to adjust for the typical magnitude difference for galaxies between $K_s$ and the UKIDSS $K-$band filter.  }
\end{center}
\end{figure} 

As a fundamental check that we indeed have succeeded in generating a magnitude limited sample of galaxies that is neither seriously contaminated by remaining stars nor plagued by the unintentional removal of a large number of galaxies, we turn to the galaxy counts as a function of apparent magnitude.  We compare our counts with those of \citet{Keen10b}, who recently combined their deep wide-field NIR counts with data from the literature to come up with the best current estimate of average galaxy counts over a wide range in apparent magnitude.  

The results of this comparison are shown in Figure~\ref{counts}.  Black data points show the counts in our sample before the aperture clipping correction was applied.  The aperture corrected counts (blue) agree well with those from \citet{Keen10b}.  These results suggest that our area estimation is accurate, our star-galaxy separation is robust, and our aperture correction method is appropriate.

\subsection{The 2M++ Catalog}
At low redshifts $(z<0.05)$, our UKIDSS sample described above is sampling a relatively small volume of the local universe.  Thus, we wish to compare our results derived from the UKIDSS sample with a low-redshift $K-$selected sample drawn from other large surveys. \citet{Lava11} published a $K_s-$selected catalog of galaxies taken from the 2MASS-XSC and cross-matched to redshifts from the 2MRS, the 6DFGRS, and the SDSS.  The resulting ``2M++" catalog is $\sim 98\%$ complete for a selection of $26,714$ galaxies over $37,080$ square degrees ($\sim 90\%$ of the sky) to a limiting magnitude of $K_{\rm{s,AB}} < 13.36$.  In the analysis that follows, we derive $K_s-$band LFs using the 2M++ to calculate the luminosity density in the local universe and compare with our results derived from the UKIDSS sample.  We also calculate the luminosity density in the $K_{\rm{s,AB}}<14.36$ subsample of the 2M++ catalog that is highly complete to somewhat higher redshifts in the SDSS and 6DFGRS regions.

\section{The $K-$band Galaxy Luminosity Function}
\label{klf}
A number of different methods exist for estimating the galaxy LF.  For an excellent review of the subject, we refer the reader to \citet{John11}.  The most commonly assumed form of the LF is that of the \citet{Sche76} function 

\begin{equation}
\Phi (L)dL = \phi^*\bigg( \frac{L}{L_*}\bigg)^\alpha {\rm exp} \bigg( \frac{-L}{L_*} \bigg) \frac{dL}{L_*},
\end{equation}

\noindent which may be written in terms of absolute magnitudes using

\begin{equation}
\frac{L}{L_*} = 10^{-0.4(M-M^*)},
\end{equation}

\noindent giving

\begin{equation}
\Phi(M) = 0.4\rm{ln}(10)\phi^* \frac{\bigg(10^{0.4(M^*-M)}\bigg)^{(\alpha+1)}}{\rm{exp}({10^{0.4(M^*-M)})}}.
\end{equation}

The Schechter function parameter $L_*$ (or $M^*$) represents the luminosity of galaxies at the knee of the LF.  $\phi^*$ determines the number density of $L_*$ galaxies, and $\alpha$ is the faint-end slope.   While the Schechter function has been demonstrated to provide a less-than-perfect fit to real data (e.g., \citealt{Jone06}), it can provide a reasonably good fit and is the most widely used functional form for fitting the LF, making it the most useful form to consider when comparing with other studies from the literature.  

\subsection{Determination of Absolute Magnitudes}
\label{absmag}
Ultimately, we wish to make a comparison of the rest-frame NIR luminosity density as a function of distance.  To do this, we need to adjust the observed apparent magnitudes in our sample by a distance modulus (DM), a $K(z)-$correction to correct for bandpass shifting, and an evolution correction, $E(z)$, such that the absolute magnitudes used for constructing luminosity functions are given as: 
\begin{equation}
M = m - DM(z) - K(z) + E(z).  
\end{equation}
At low redshifts in the NIR, $K(z)-$corrections are nearly independent of galaxy type \citep{Mann01}, allowing this magnitude correction to be made without considering the type distribution of the galaxy sample.  Combining SDSS and UKIDSS data, \citet{Chil10} showed that, at $z<0.5$, accurate $K-$corrections can be calculated using low-order polynomials with input parameters of only the redshift and one observed NIR color.  They have provided a $K-$correction calculator package\footnote{http://kcor.sai.msu.ru/}, which we used to compute $K-$corrections for galaxies in our sample.  \citet{Chil10} compared their $K-$correction calculator output values with those obtained via the more rigorous spectral energy distribution fitting methods of \citet{Blan07} and \citet{Fioc97} and concluded that the magnitude errors associated with $K-$corrections derived using their algorithm should be $< 0.1$ magnitudes.  

Evolution of the rest-frame NIR light from galaxies is expected to be significantly weaker than in optical bandpasses \citep{Blan03}, but it is an effect that must be accounted for when comparing galaxy luminosities at different redshifts.  A commonly assumed form of the evolution correction is $E(z) = Qz$, where $Q$ is a positive constant.  \citet{Blan03} showed that in the NIR, $Q=1$ agrees well with  stellar population synthesis models.  Thus, for this study, we adopt $Q=1$, such that $E(z) = z$.  We further discuss this evolution correction, and its associated uncertainties, in Section~\ref{lfshape}.

\subsection{Fitting Methods}
To fit Schechter functions to observed data, a variety of methods have been used in the past.  In \citet{Keen12}, we compared four different LF estimators ($1/V_{\rm{max}}, C^-$, STY, and SWML) in the determination of NIR LFs.  We found that the STY \citep{Sand79} and SWML (Step-Wise Maximum Likelihood, \citealt{Efst88}) methods yielded similar results in the determination of $M^*$ and $\alpha$, while the $C^-$ \citep{Lynd71} and $1/V_{\rm{max}}$ \citep{Schm68} methods tended to underestimate the faint end slope (see also \citealt{Page2000}).  Of these four methods, $1/V_{\rm{max}}$ is the only one that provides the normalization ($\phi^*$).  In \citet{Keen12}, we tested the $1/V_{\rm{max}}$ method alongside three other normalization estimators from \citet{Davi82}.   We found all four of these estimators  yielded consistent results, a confirmation of the same result found by \citet{Will97} using simulated data.  

Given these analyses, in \citet{Keen12}, we settled on a hybrid method to estimate the LF by first using STY to calculate $M^*$ and $\alpha$, and then using $1/V_{\rm{max}}$ (with $M^*$ and $\alpha$ fixed) to determine the normalization.  Here we use this same hybrid method in the determination of the $K-$band LF.   To correct for spectroscopic incompleteness, we use a simple scheme  in which each galaxy counted in the $1/V_{\rm{max}}$ procedure is weighted by a factor of $1/C(m)$, where $C(m)$ is the fractional completeness as a function of apparent magnitude.

\subsection{The Assumption of a Constant LF Shape}
\label{constantlf}
 We assume that the $z=0$ shape parameters of the $K-$band LF ($M^*$ and $\alpha$) are not changing as a function of environment or distance from us.  We require this assumption to facilitate the measurement of the $K-$band luminosity density as a function of distance.  This is due to the fact that, in this sample, we have limited information about the faint end slope of the LF in more distant volumes (higher redshifts) due to the magnitude limit of the survey, and we have corrupted information about the bright end of the LF of nearby galaxies  (lower redshifts) due to the Petrosian aperture clipping issue described in Section~\ref{clipping}.  

We believe the assumption of a constant LF shape is reasonable, given that the $K(z)-$corrections are essentially independent of galaxy type and the $E(z)$ corrections are quite modest (see Section~\ref{absmag}).    Furthermore, \citet{Depr09} and \citet{Capo12} have found that the shape of the NIR LF is not significantly different for field and cluster galaxies.  

\subsection{The $K-$band luminosity density at $z < 0.2$}
\label{kbandlf}
\begin{figure}
\begin{center}
\includegraphics[width=90mm]{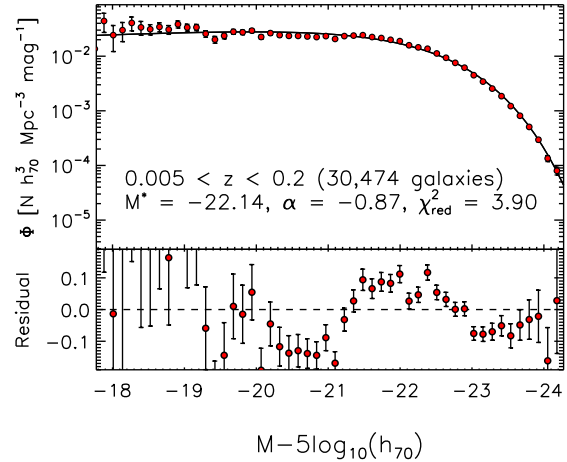}
\caption{\label{allklf} (a) The UKIDSS-LAS $K-$band luminosity function over the range $0.005<z<0.2$.  The red circles show the $1/V_{\rm{max}}$ LF estimate for this redshift range.  Error bars show Poisson counting errors.  The curve shows the best fit Schechter function to the $1/V_{\rm{max}}$ data after having determined the values of $M^*=-21.37$ and $\alpha=-0.87$ using the STY method.   The value of $\chi^2_{\rm{red}}\sim 4$ demonstrates the relatively poor fit of the Schechter function to the data.  The strange shaped residuals could be due, in principle, to the Schechter function being a poor model of the data, or inhomogeneity along the line of sight. }
\end{center}
\end{figure} 

\begin{figure}
\begin{center}
\includegraphics[width=85mm]{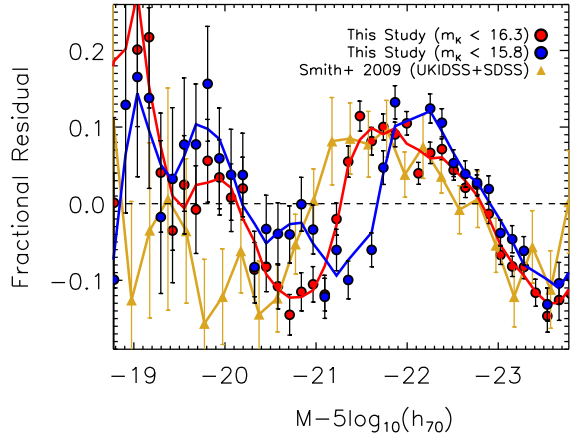}
\caption{\label{residuals} Residuals for this study at two different magnitude limits ($K_{\rm{AB}}<16.3$ in red and $K_{\rm{AB}}<15.8$ in blue).  The fact that the whole feature moves to brighter absolute magnitudes (blue vs. red above) upon changing the magnitude limit of the sample suggests this feature is due to inhomogeneity along the line of sight and not intrinsic to the LF.  The sample considered by \citet{Smit09} (residuals shown in yellow) is largely overlapping with our sample, but to a slightly fainter $K-$band limit, and indeed they observe the same feature in the residuals shifted slightly to fainter absolute magnitudes.}
\end{center}
\end{figure} 

\begin{figure}
\begin{center}
\includegraphics[width=90mm]{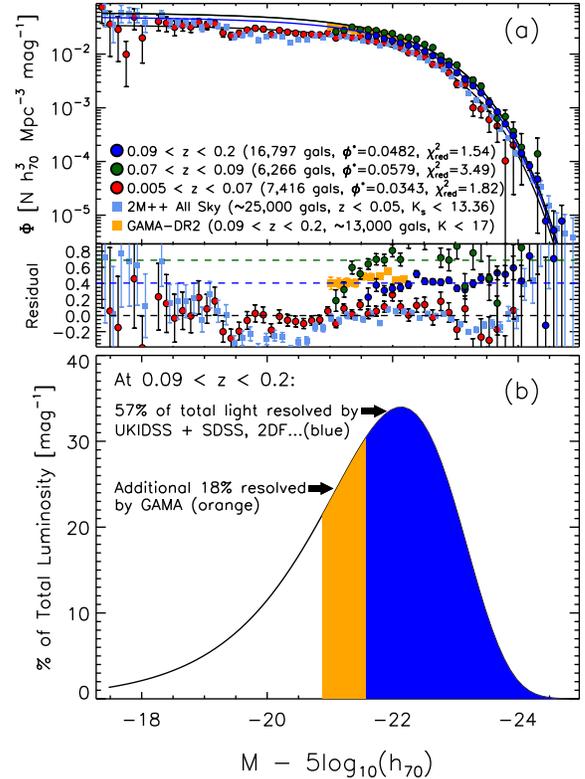}
\caption{\label{lovhi} (a) The UKIDSS-LAS $K-$band luminosity function split into three redshift ranges: $0.005<z<0.07$ (red), $0.07<z<0.09$ (green), and $0.09< z < 0.2$ (blue).  We separate the redshift range $0.07<z<0.09$ to demonstrate that the excess at $z > 0.09$ is not due to the Sloan Great Wall or the other over-densities we observe at higher declination in this redshift range.    Here we fit $\phi^*$ in each redshift range with $\alpha$ and $M^*$ fixed (as described in Section~\ref{kbandlf}).  We list the $\chi^2_{\rm{red}}$ values for each redshift range in the figure, and note that relatively good fits can be obtained using fixed LF shape parameters and letting the normalization ($\phi^*$) vary as a function of redshift.   Residuals are shown relative to the low-redshift normalization.  Here we also include our own analysis of the 2M++ catalog (all sky, $K_{\rm{s,AB}}<13.36,~\sim 25,000$ galaxies).  We have fit the normalization of the LF derived from the 2M++ catalog (LF and residuals shown as light blue squares) with the same shape parameters as for the UKIDSS sample.   To extend the faint end of the $z>0.09$ LF, we use the GAMA redshift sample ($K_{\rm{AB}}<17$, shown as orange squares).  (b) The relative contribution to the total luminosity density (\% per magnitude) as a function of absolute magnitude.  The blue shaded region shows the range of magnitudes covered by the UKIDSS sample at $z>0.09$.  In orange we show the extra fraction of light resolved by extending the faint end of the LF with the GAMA sample.  This demonstrates that, at $z>0.09$, we are making a robust measurement of the peak of the luminosity density distribution (occurring at $M\sim M^*$) and resolving $\sim 75\%$ of the total light.}
\end{center}
\end{figure} 

In Figure~\ref{allklf}, we show the UKIDSS $K-$band LF for all galaxies in the redshift range $0.005 < z < 0.2$.  The red circles show the $1/V_{\rm{max}}$ estimate of the LF, and the curve shows a fit to these data having determined $M^*$ and $\alpha$ using STY and then fitting $\phi^*$ to the $1/V_{\rm{max}}$ data.  We note the poor fit to the data ($\chi^2_{\rm{red}} \sim 4$).   A poor fit of the Schechter function to LF estimates is not uncommon in studies from the literature (see e.g., \citealt{Lava11, Smit09, Jone06}), and authors normally attribute this to the Schechter function being a poor model of the data.  In the analysis that follows, however, we demonstrate that much of this fitting error, and the shape of the residuals, may be explained by a rising space density of galaxies along the line of sight.  

In Figure~\ref{residuals}, we show the residuals for the UKIDSS sample at two different flux limits for the sample ($K_{\rm{AB}}<16.3$ in red and $K_{\rm{AB}}<15.8$ in blue).  We note that the whole sinusoidal feature in the residuals appears to move $\sim 0.5$ magnitudes to the right with a change of $0.5$ mags in the flux limit of the sample.  This indicates that this feature in the residuals is not intrinsic to the LF, but due to inhomogeneity along the line of sight.  We also compare with the residuals from \citet{Smit09} (shown in yellow) for their largely overlapping sample with a slightly fainter flux limit.  Indeed, they see the same feature shifted slightly to fainter absolute magnitudes, as expected if this feature is not intrinsic to the LF, but rather due to large-scale structure in the sample.

We now divide our sample into redshift bins to further consider the matter of inhomogeneity along the line of sight.  It is worth noting at this point that a considerable portion $(\sim25\%)$ of the solid angle  subtended on the sky by this sample is filled by the Sloan Great Wall (\citealt{Gott05}) at redshifts of $0.07<z<0.09$.  In our sample, $\sim 40\%$ of the galaxies in this redshift range are part of this structure.  The relative excess in the redshift distribution due to the Sloan Great Wall can be seen as a peak at these redshifts in Figure~\ref{zhist}, both in the overall redshift distribution and, more prominently, in subregion 2, which is centered on this structure.    Such structures will certainly be expected to cause some deformity in the LF, given that different luminosity ranges are preferentially sampled at different redshifts in an apparent magnitude limited survey.

In Figure~\ref{lovhi}a, we show the LF for galaxies in the range $0.005<z<0.07$ as red circles, for $0.07<z<0.09$ as green circles, and for $0.09<z<0.2$ as blue circles.  The LF shape parameters, $M^*$ and $\alpha$, are the same for all three LFs shown in this figure.  We determined $M^*$ and $\alpha$ iteratively using the STY method by fitting $M^*$ on the high-redshift ($0.09<z<0.2$) sample with $\alpha$ fixed, then fitting $\alpha$ on the low-redshift $(0.005 < z < 0.07$) sample with $M^*$ fixed.  We repeated this procedure until we reached convergence at values of $M^* = -22.15\pm 0.04$ and $\alpha=-1.02\pm 0.03$.  

With $M^*$ and $\alpha$ fixed, we fit the normalization, $\phi^*$, to the LF estimate given by the $1/V_{\rm{max}}$ method.  We find that the shape of the LFs for the high-redshift and low-redshift samples shown in Figure~\ref{lovhi} can be reasonably well fit ($\chi^2_{\rm{red}} \sim 1$ and relatively flat residuals, compared to Figure~\ref{allklf}) by the same $M^*$ and $\alpha$ parameters, while only allowing the normalization to change (we attribute the comparably poor fit at $0.07<z<0.09$ to the large inhomogeneities in this redshift bin).  This indicates that the assumption of a constant LF shape as a function of redshift and environment appears to be valid.  As shown in Figure~\ref{lovhi}, we find a normalization at $0.09<z<0.2$ that is $\sim 1.4$ times higher than that at $0.005 < z<0.07$.  Thus, if the LF shape is indeed constant for this sample, then the luminosity density at $z>0.09$ appears to be $\sim 1.4$ times higher than that at $z<0.07$, even when known over-densities at  $0.07<z<0.09$ are excluded.  

We note that the relatively small volume ($10^6$ Mpc$^3$) probed in the UKIDSS sample at $z<0.07$ could be considered too small to be representative, so we turn to the 2M++ sample for a better measure of the local $K-$band luminosity density.  The same quantities (LF and residual) are shown in panel (a) of Figure~\ref{lovhi} for the 2M++ sample as small light blue squares.  This sample is $\sim 98\%$ complete over $\sim 90\%$ of the sky to $K_{\rm{s,AB}}<13.36$ and consists of $\sim 25,000$ galaxies.  Here we have shifted the 2M++ $K_s-$band LF by $+0.15$ mags to account for the typical offset between UKIDSS $K$ and 2MASS $K_s$ magnitudes.

We derived the 2M++ LF shown in Figure~\ref{lovhi} in the same way as that for the UKIDSS sample for $0.001<z<0.05$, and we simply fit the normalization using the same values for $M^*$ and $\alpha$ as derived from the UKIDSS data.  We find the UKIDSS and 2M++ LFs agree very well at absolute magnitudes of $M_K<-21.25$.  Given the shape of the LF, roughly $\sim 80\%$ of the total luminosity density is coming from this magnitude range, so we take this as evidence that our low-redshift measurement of the $K-$band LF using the UKIDSS data is accurate.  Furthermore, our estimates of the total luminosity density made by integrating either of the UKIDSS or 2M++ LFs presented here are in good agreement with both the estimates of the same quantity made by \citet{Lava11} using independent methods and those of \citet{Bran12} using a similar sample.

Our measurement of the LF at $0.09<z<0.2$ is made over a similar size volume ($\sim 10^7$ [Mpc/h]$^3$) to that of the 2M++ catalog at low redshifts.  However, we only measure the LF of galaxies at $\sim M_K<-20.77$ in the higher redshift UKIDSS sample.  The GAMA-DR2 sample features independent $K-$band photometry from the UKIDSS-LAS and highly complete ($\sim 98\%$) spectroscopy at $K_{\rm{AB}}<17$ over $144$ deg$^2$ on the sky.  Thus, we use the GAMA-DR2 sample  to extend our measurement of $z>0.09$ LF to $\sim 0.7$ magnitudes fainter.  In Figure~\ref{lovhi}, we show the LF and residual that we derive for the GAMA sample in orange.  We find that the $z>0.09$ LF from the GAMA sample appears to be a relatively good match to the $z>0.09$ UKIDSS LF and provides further evidence that the shape of the LF is the same in the local and more distant volumes.  

In Figure~\ref{lovhi}b, we show the percent contribution (in percent per magnitude) to the total luminosity density as a function of absolute magnitude (for a Schechter function with $M^*=-22.15$ and $\alpha=-1.02$).  The peak of the luminosity density distribution is at $M\sim M^*$.  The area shaded in blue shows the fraction resolved by the UKIDSS sample ($\sim 57\%$ of the total light).  The area shaded in orange shows the contribution of the GAMA sample resolving an additional $\sim 18\%$ of the total light.   

Thus, the strange looking residuals seen in Figure~\ref{allklf} may be deconstructed into a low-redshift and high-redshift component.   They are naturally explained by a LF that takes a shape quite similar to a Schechter function with a higher space density of galaxies at higher redshifts.  While we are only measuring the LF over a limited range in absolute magnitude at $z>0.09$ in our sample ($M_K<-21$), this corresponds to a robust measurement of the peak of the luminosity density distribution and $\sim 75\%$ of the total light (as shown in Figure~\ref{lovhi}).  

\subsection{A Rising $K-$band Luminosity Density as a Function of Distance}

\begin{figure*}
\begin{center}
\includegraphics[width=170mm]{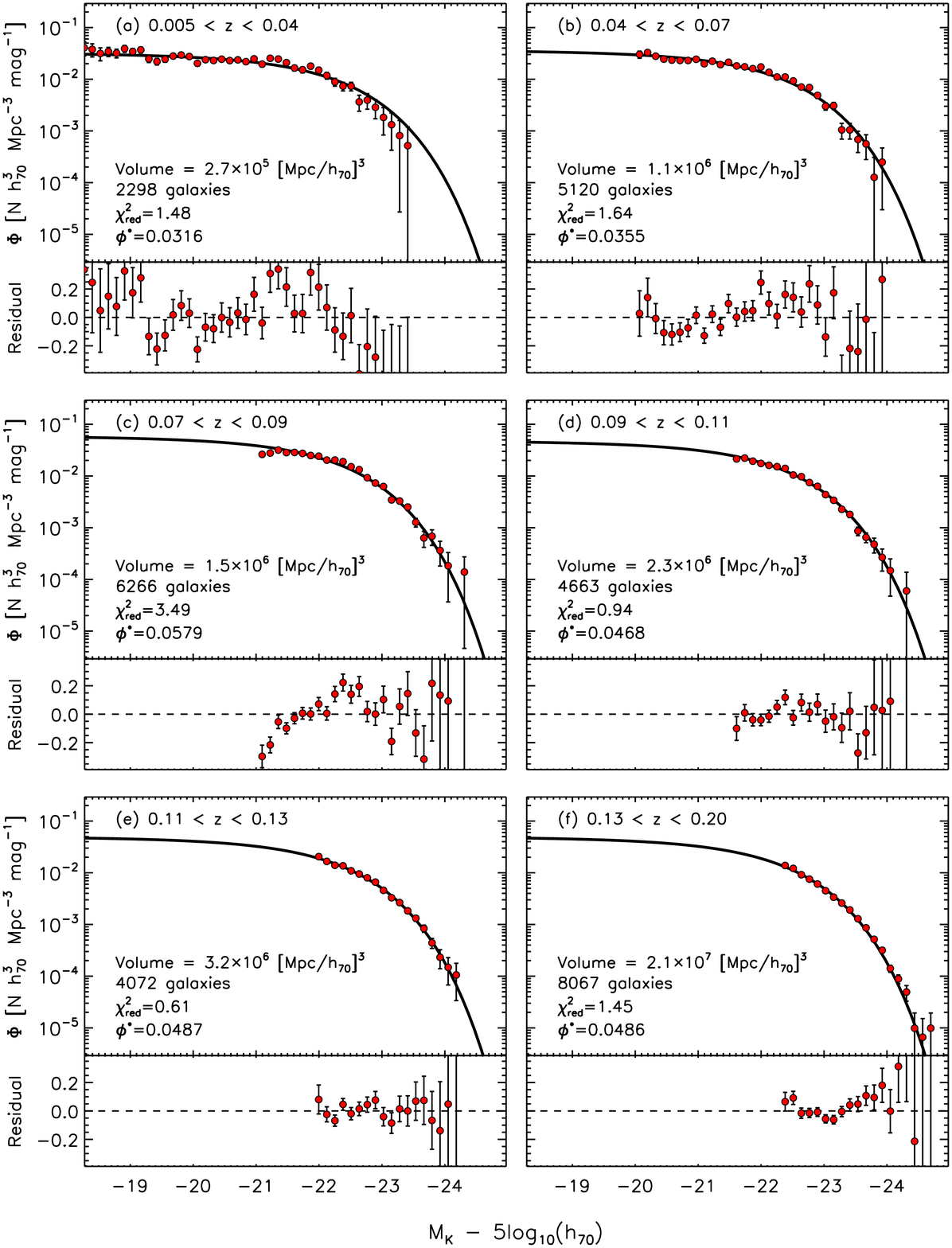}
\caption{\label{uklf} UKIDSS-LAS $K-$band LFs as a function of redshift.   Here we have fixed the LF shape parameters ($\alpha = -1.02$ and $M^* = -22.15$) to those derived from the iterative fitting method described in Section~\ref{constantlf} using the data presented in  Figure~\ref{lovhi}.  We then divided the sample into the six independent redshift bins shown here.  In each case, the error bars show Poisson counting errors.  Residuals are shown in the bottom panel of each plot.  We find that we can get a relatively good fit over all redshift bins (with the notable exception of the $0.07<z<0.09$ bin containing the Sloan Great Wall) by simply letting the Schechter function normalization vary as a function of redshift.  The volume and number of galaxies used to determine the LF in each bin are listed in the plots, as well as the $\chi^2_{\rm{red}}$ values and normalization ($\phi^*$) for each case.}
\end{center}
\end{figure*}

Next, we investigated the $K-$band luminosity density as a function of distance in a series of narrower redshift bins over the range $0.005<z<0.2$, as labeled in each panel of Figure~\ref{uklf}.  We fixed the LF shape parameters to the values derived above ($\alpha=-1.02$ and $M^*=-22.15$) and fit the normalization to the $1/V_{\rm{max}}$ results in each redshift bin.  Red circles show the $1/V_{\rm{max}}$ results in each case, and the black solid curves represent the Schechter function fits.  Again, we find that the data in all of the redshift bins are reasonably well fit with the same $M^*$ and $\alpha$ parameters ($\chi^2_{\rm{red}} \sim 1$, except at $0.07<z<0.09$), suggesting that the assumption of a constant LF shape is valid.  For each redshift slice, we list in the plots the volume sampled, the number of galaxies used to generate the LF, the $\chi^2_{\rm{red}}$ value for the fit, and the value of $\phi^*$.

\begin{figure*}
\begin{center}
\includegraphics[width=150mm]{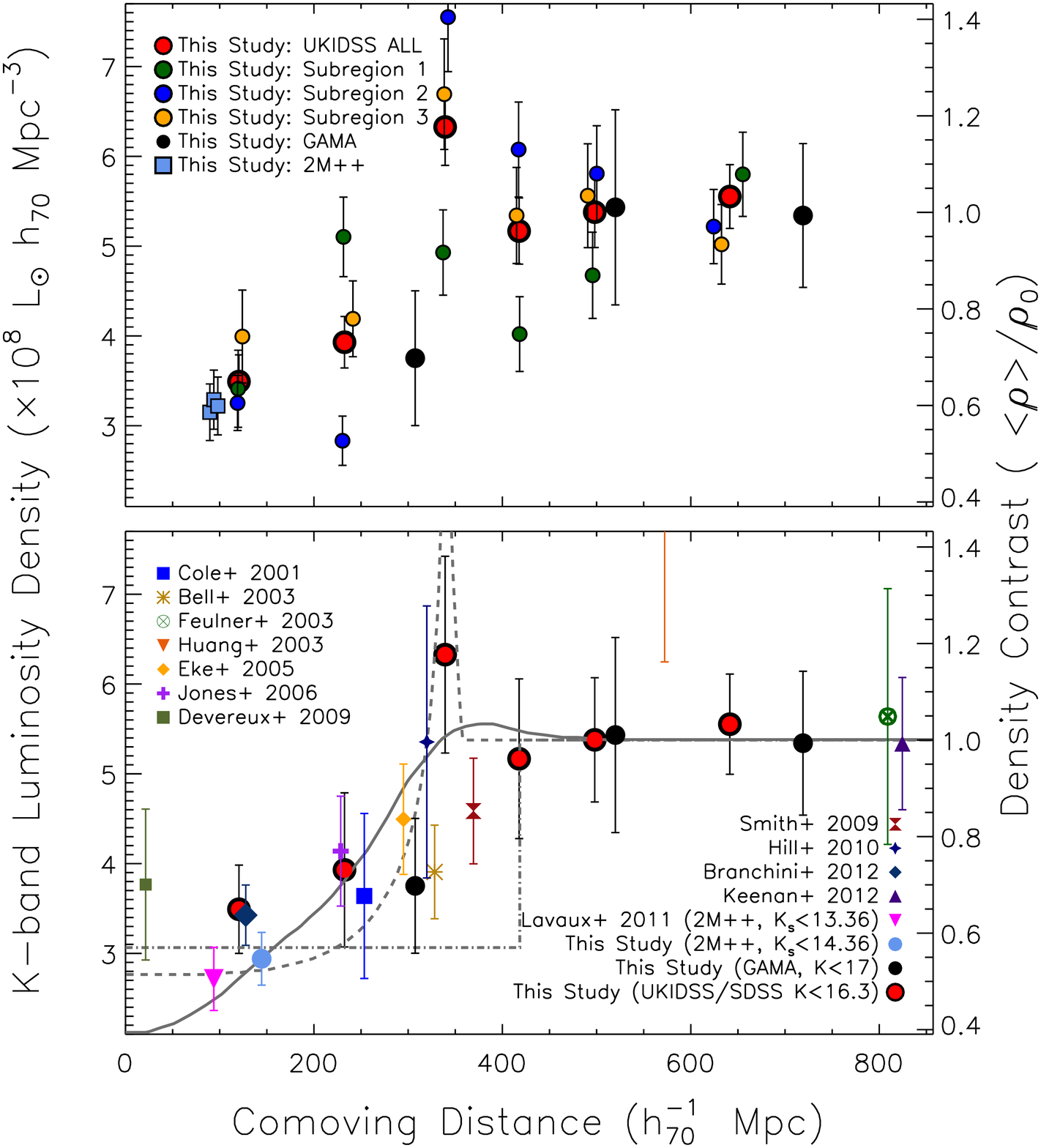}
\caption{\label{ukld} $K-$band luminosity density as a function of comoving distance.  (a) Our measured $K-$band luminosity density for the full sample (red circles) versus different directions on the sky (green, blue, and orange circles corresponding to subregions 1, 2, and 3, respectively, in Figure~\ref{cover}). Light blue squares indicate the $K-$band luminosity density we measure in three different directions (SDSS, 6DFGRS, and 2MR regions) using the 2M++ all sky catalog compiled by \citet{Lava11}.    (b) Our measured $K-$band luminosity density for the full sample (red circles) as a function of comoving distance compared with other studies from the literature.   Our estimate of the $K_{\rm{s,AB}}<14.36$ luminosity density from the 2M++ catalog (SDSS and 6DFGRS regions only) is shown as a light blue circle.  Our estimates in three redshift bins for the GAMA survey only ($K_{\rm{AB}}<17$, same methods as for the UKIDSS sample) are shown as black circles.  All of the data shown in this plot are listed in Table~\ref{table1}.  The density contrast, $\langle \rho \rangle / \rho_0$, is displayed on the right-hand vertical axis.  The scale of the right-hand axis was established by performing an error-weighted least-squares fit (for the normalization only, not shape) of the radial density profile of \citet{Bole11a} (gray solid curve) to all the luminosity density data in panel (b).  The dashed curve shows the radial density profile of \citet{Alex09}.  Both \citet{Alex09} and \citet{Bole11a} claimed these density profiles can provide for good fits to the SNIa data without dark energy.   The dash-dot curve shows the scale and amplitude of the ``Hubble bubble" type perturbation that \citet{Marr13} would require to explain the discrepancy between local measurements of the Hubble constant and those inferred by Planck.}
\end{center}
\end{figure*}

\begin{deluxetable*}{lcccccccc}
\tabletypesize{\tiny}
\tablewidth{0pt}
\tablecaption{\label{table1}NIR luminosity densities and Schechter function parameters for this study and selections from the literature}
\tablehead{ &$\langle z \rangle$\tablenotemark{a}  & Redshift Range & N$_{gals}$ & $M^*$\tablenotemark{b} & $\alpha$& $\phi^*$\tablenotemark{c}$\times 10^3$ & $j_{\rm K, calc}$\tablenotemark{d} & $j_{\rm K,pub}$\tablenotemark{e}}
\startdata
This Study (UKIDSS) & $0.028$ & $0.005-0.04$ & $2,298$ & $-22.15\pm0.04$ & $-1.02\pm0.03$ & $31.6\pm1.5$ & $3.58\pm 0.50$ & $-$  \\
This Study (UKIDSS)& $0.055$ & $0.04-0.70$ & $5,120$ & $-22.15\pm0.04$ & $-1.02\pm0.03$ & $35.5\pm1.7$ & $4.02\pm0.87$ & $-$ \\
This Study (UKIDSS)& $0.081$ & $0.07-0.09$ & $6,266$ & $-22.15\pm0.04$ & $-1.02\pm0.03$ & $57.9\pm2.3$& $6.55\pm1.12$ & $-$ \\
This Study (UKIDSS)& $0.099$ & $0.09-0.11$ & $4,663$ & $-22.15\pm0.04$ & $-1.02\pm0.03$ & $46.8\pm1.2$ & $5.30\pm 0.90$ & $-$ \\
This Study (UKIDSS)& $0.119$ & $0.11-0.13$ & $4,072$ & $-22.15\pm0.04$ & $-1.02\pm0.03$ & $48.7\pm1.5$ & $5.51 \pm 0.71$ & $-$ \\
This Study (UKIDSS)& $0.154$ & $0.13-0.20$ & $8,067$ & $-22.15\pm0.04$ & $-1.02\pm0.03$ & $48.6\pm1.2$ & $5.50\pm0.56$ & $-$ \\
This Study (2M++)& $0.034$ & $0.01-0.07$ & $53,000$ & $-22.15\pm0.04$ & $-1.02\pm0.03$ & $29.7\pm2.9$& $2.95\pm0.21$ & $-$ \\
This Study (GAMA)& $0.073$ & $0.05-0.10$ & $4,108$ & $-22.15\pm0.04$ & $-1.02\pm0.03$ & $33.5\pm1.1$ & $3.75\pm 0.90$ & $-$ \\
This Study (GAMA)& $0.125$ & $0.10-0.15$ & $6,790$ & $-22.15\pm0.04$ & $-1.02\pm0.03$ & $48.7\pm1.6$ & $5.43 \pm 0.71$ & $-$ \\
This Study (GAMA)& $0.175$ & $0.15-0.20$ & $5,488$ & $-22.15\pm0.04$ & $-1.02\pm0.03$ & $47.2\pm1.2$ & $5.34\pm0.56$ & $-$ \\
Keenan+ (2012)& $0.200$ & $0.10-0.30$ & $812$ & $-22.33 \pm 0.06$ &  $-0.91 \pm 0.07$ & $42.6 \pm 4.9$ & $5.23 \pm 0.72$ &  $-$ \\

Branchini+ (2012)& $0.030$ & $0.001-0.08$ & $45,000$ & $-22.46 \pm 0.03$ &  $-1.00 \pm 0.02$ & $26.7 \pm 3.8$ & $3.42 \pm 0.34$ &  $-$ \\

Lavaux+ (2011)& $0.022$ & $0.0025-0.067$ & $60,000$ & $-22.10 \pm 0.02$ &  $-0.73 \pm 0.02$ & $32.4 \pm 0.6$ & $2.67 \pm 0.18 $ & ($2.76 \pm 0.02$)  \\

Hill+ (2010) & $0.076$ & $0.003-0.1$ & $1,785$ & $-22.43 \pm 0.10$ &  $-0.96 \pm 0.06$ & $45.5 \pm 4.7$ & $5.36 \pm 1.51$ & ($4.89 \pm 1.13$)  \\

Smith+ (2009) & $0.100$ & $0.01-0.3$ & $40,111$ & $-22.25 \pm 0.04$ &  $-0.81 \pm 0.04$ & $48.4 \pm 2.3$ & $4.59 \pm 0.59$ &  ($4.33 \pm 0.05$)\\

Devereux+ (2009) & $0.005$ & $0.001-0.01$ & $1,349$ & $-22.33 \pm 0.46$ &  $-0.94 \pm 0.10$ & $33.5 \pm 9.9$ & $3.77 \pm 0.84$ &  ($4.06 \pm 0.84$)\\

Jones+ (2006) & $0.054$ & $0.0025-0.15$ & $60,869$ & $-22.75 \pm 0.03$ &  $-1.16 \pm 0.04$ & $21.9 \pm 1.5$ & $4.06 \pm 0.60$ & ($4.06 \pm 0.35$)  \\

Eke+ (2005) & $0.070$ & $0.005-0.12$ & $15,644$ & $-22.35 \pm 0.04$ &  $-0.81 \pm 0.07$ & $41.7 \pm 2.3$ & $4.41 \pm 0.60$ & ($4.93 \pm 0.16$)   \\

Bell+ (2003) & $0.078$ & $0.0033-0.2$ & $6,282$ & $-22.21 \pm 0.05$ &  $-0.77 \pm 0.04$ & $41.7 \pm 2.0$ & $3.84 \pm 0.51$ & ($4.06 \pm 1.26$) \\

Huang+ (2003) & $0.138$ & $0.005-0.35$ & $1,056$ & $-22.62 \pm 0.08$ &  $-1.37 \pm 0.10$ & $37.9 \pm 8.7$ & $7.92 \pm 1.79$ & $-$  \\

Feulner+ (2003) & $0.200$ & $0.10-0.30$ & $210$ & $-22.71 \pm 0.24$ &  $-1.10 \pm 0.10$ & $32.4 \pm 3.5$ & $5.54 \pm 1.40$ & $-$  \\

Cole+ (2001) & $0.060$ & $0.023-0.12$ & $5,683$ & $-22.36 \pm 0.03$ &  $-0.96 \pm 0.06$ & $31.5 \pm 4.7$ & $3.57 \pm 0.90$ & ($4.02 \pm 0.60$)  \\

\enddata
\tablenotetext{a}{The mean redshift values listed for this study are averages for each redshift bin.  In other studies the value in this column is not strictly a mean, but rather, whatever the authors listed (mean or median) or our own estimate, if no value was given in the original article.}
\tablenotetext{b}{$M^*$ values are given as $M-5$log$_{10}(h_{70})$.  Note: these are the published values; however, in calculating the luminosity density, we adjusted studies using UKIDSS $K-$band Petrosian magnitudes (this study, \citealt{Smit09}, \citealt{Hill10}) to 0.15 mags brighter in $M^*$ for consistency with 2MASS ``total" $K_s-$band magnitudes.}
\tablenotetext{c}{$\phi^*$ values are given in units of $h_{70}^3$~Mpc$^{-3}$.}
\tablenotetext{d}{$K-$band luminosity density in units of $10^8 L_{\odot}h_{70}$~Mpc$^{-3}$.  These are the values used in Figure~\ref{ukld} for which we have calculated $j_{\rm K}$ such that all studies are assuming the same value for the solar luminosity in the $K-$band of $M_{\odot,\rm K}=5.14$ and the same limits of integration of the Schechter function ($M^*-5$ to $M^*+10$).  The errors listed for this study are statistical plus an estimate of the systematic uncertainty due to cosmic variance.}
\tablenotetext{e}{$K-$band luminosity density as published in the studies listed.  These values differ from those we calculate primarily due to different assumptions about the value of $M_{\odot,\rm K}$.}
\end{deluxetable*}

In Figure~\ref{ukld}, we use the results from Figure~\ref{uklf} (and similar results for the three different subregions shown in Figure~\ref{cover} individually) to calculate the $K-$band luminosity density as a function of distance (by integrating the fitted Schechter functions).  In Figure~\ref{ukld}a we show how these results vary as a function of position on the sky by comparing the average result for the full sample (red circles) with the results from each of the three subsamples (green, blue, and orange circles) of galaxies from the subregions shown in Figure~\ref{cover}.  Each of these subsamples contains roughly a third of the original sample of galaxies.  While we find stark differences between subsamples in the measured luminosity density in some redshift bins, the overall trend toward a rising luminosity density with increasing redshift appears present in all cases.  We use the rms variation in luminosity density between the subregions in each redshift bin as an estimate of the systematic error due to cosmic variance in our measurement.  These systematics are reflected in the larger error bars for the total sample in Figure~\ref{ukld}b.  

In Figure~\ref{ukld}a, we also present a reanalysis of the 2M++ catalog, where we have used the same techniques to measure the luminosity density as we applied to the UKIDSS data.  We have divided the 2M++ sample up into 3 independent subsamples as a probe of cosmic variance and to verify our low-redshift results that were made over a relatively small volume.  The three light blue squares in Figure~\ref{ukld}a represent our measured luminosity density using the 2M++ catalog in the SDSS region ($\sim 7,500$ deg$^2$), the 6DFGRS region ($\sim 17,000$ deg$^2$), and the 2MR region ($\sim 12,500$ deg$^2$).  These results agree well with one another and with our luminosity density measurements from UKIDSS data.  Furthermore, these results agree well with an independent analysis of the 2M++ catalog by \citet{Lava11} and analysis of a similar dataset by \citet{Bran12}.  Our estimates in three redshift bins using the GAMA-DR2 survey data alone ($K_{\rm{AB}}<17$) are shown as black circles.    

In Figure~\ref{ukld}b, we compare our results with other studies from the literature, where red circles again show our results for the entire sample, and studies from the literature correspond to the symbols listed in the plot legend.  We also show our own estimate of the luminosity density for the 2M++ $K_{\rm{s,AB}}<14.36$ sample as a light blue circle.  The luminosity densities for the literature studies have been recalculated here using the same value for the solar luminosity in the $K-$band of $M_{\odot,\rm{K}}=5.14$.  This recalculation does not result in substantially different values from those published in the original studies, but we do it for consistency.  All the data displayed here are listed in Table~\ref{table1}.  

The majority of studies at $z<0.1$ have used photometry from the Two Micron All Sky Survey (2MASS, \citealt{Jarr00}).   These studies have typically used the 2MASS $K_s-$band ($2.12\mu$m) Kron or ``total" magnitudes, which are generally found to be $\sim 0.15$ mags brighter than UKIDSS $K-$band Petrosian magnitudes (see e.g., \citealt{Keen10a, Smit09}).  Thus, in making this comparison, we have adjusted the values for $M^*$ to be $0.15$ mags brighter for our study and for other studies derived from UKIDSS data \citep{Hill10,Smit09}.  Making this adjustment does not change the results presented here, but we do it for the sake of consistency.    In general,  we find excellent agreement with all previously published results in the $K-$band, and we confirm the tentative result presented in \citet{Keen12} of a rising luminosity density from distances of $250-350$ Mpc that appears to remain higher than that measured locally out to $D\sim 800$ Mpc. 

The relative density contrast, $\langle \rho \rangle / \rho_0$, is displayed on the right-hand vertical axis in Figure~\ref{ukld}.  The scale of this axis was established by performing an error-weighted least-squares fit of the radial density profile of \citet{Bole11a} (solid curve, fitting the normalization only, not the shape) to all the luminosity density data in panel (b).   As mentioned in Section~\ref{intro}, this means we are assuming a linear bias parameter of $b=1$, or, more specifically, that $K-$band luminosity density is an unbiased tracer of the underlying dark matter density.  We believe this is a reasonable assumption given previous results from observations and simulations (e.g., \citealt{Mall05,Angu13}).  

The dashed curve in Figure~\ref{ukld} shows the radial density profile of \citet{Alex09}, also normalized to a density contrast of unity at high redshift.  Both \citet{Alex09} and \citet{Bole11a} claim these density profiles allow for a good fit to the SNIa data without dark energy.  The dash-dot curve shows the scale and amplitude of the ``Hubble bubble" type perturbation that \citet{Marr13} would  require to explain the discrepancy between local measurements of the Hubble constant and those inferred by Planck.   

Although we show the low-redshift study performed by \citet{Deve09}, their luminosity density is biased strongly by the very local over-density known as the Supergalactic Plane (containing roughly $\sim 40\%$ of the galaxies in the sample), so we believe this is also probably an overestimate of the average luminosity density on larger scales in the local universe.   

\citet{Hill10} use UKIDSS data to measure the $K-$band luminosity density in the field targeted by the Millennium Galaxy Catalog Survey (${\rm Dec}=0, 10^h < {\rm RA} < 15^h$, \citealt{Lisk03}).  As noted in Section~\ref{kbandlf}, this region contains the Sloan Great Wall.  \citet{Hill10} correct for this over-density by adjusting their normalization down by $\sim 20\%$.  Undoing this correction would bring their measurement into good agreement with our result at roughly the same redshift.  

The sample of \citet{Smit09} is the most similar to our own.  They select $\sim 40,000$ galaxies from the UKIDSS-LAS with redshifts from the SDSS.  They measure the LF for the entire sample from $0.01<z<0.3$.  They find similar irregularities in the LF shape as shown in Figure~\ref{residuals}.  They note that when they divide their sample into three redshift bins they see a rising LF normalization with increasing redshift, and they point out that even doubling their evolution correction does not resolve this issue.  We note that \citet{Jone06} also find very similar shaped residuals in both optical and NIR bands in their study of the LF using the 6DFGRS.  

The sample sizes of \citet{Huan03}, \citet{Feul03}, and \citet{Keen12} are generally too small to be considered robust to cosmic variance.  However, it is worth noting that the measurements of \citet{Feul03} and \citet{Keen12} may be considered conservative underestimates of the true luminosity density, because these studies avoided known over-densities, such as galaxy clusters, in the redshift ranges sampled.   

We conclude that if the observed trend in $K-$band luminosity density as a function of distance is indicative of a similar trend in the underlying  total mass density, then the local universe may be under-dense on a scale and amplitude sufficient to introduce significant biases into local measurements of cosmological observables.  Leaving aside considerations of whether or not such an unusual local structure could obviate the need for dark energy, it appears that the observed under-density is roughly the right scale and amplitude (given the analysis of \citealt{Marr13}) to explain the apparent tension between local measurements of the Hubble constant ($H_0=73.8\pm 2.4$ km~s$^{-1}$~Mpc$^{-1}$, \citealt{Ries11}) and the recent results from Planck ($H_0=67.3\pm 1.2$ km~s$^{-1}$~Mpc$^{-1}$, \citealt{Ade13}).

\section{Possible Sources of Bias and Error}
 \label{biases}
We have done everything possible in this study to make an unbiased measurement of the $K-$band luminosity density as a function of redshift in the nearby universe.  However, we are making several assumptions that allow us to probe a wider redshift range than that for which we have a statistically robust measurement of the LF over the entire range of absolute magnitudes.  Here we discuss the possible biases and errors in our measurement associated with these assumptions.

\subsection{K(z) and E(z) corrections}
\label{lfshape}
The most important assumption we make in this study is that the $z=0$ shape of the  $K-$band LF is not changing significantly as a function of distance from us.  This, in turn, relies on the assumption that we are making the appropriate $K(z)$ and $E(z)$ corrections to calculate the $z=0$ absolute magnitudes of galaxies at any given redshift.  To investigate these assumptions in detail, we explored variations in the $K(z)$ and $E(z)$ corrections to see how they affect our result.

If we simply omit the $K(z)-$corrections in our calculations, we find that the excess luminosity density at $z>0.09$ shown in Figure~\ref{lovhi} increases significantly.  This is due to the fact that the $K-$correction is negative, i.e., making this correction acts to reduce the observed flux, and the reduction is greater at higher redshifts.  We compared the $K-$correction calculator outputs of \citet{Chil10} to the empirical estimates by \citet{Mann01} and found qualitative agreement, but on average the $K-$corrections of \citet{Mann01} are $\sim 0.1$ mags more negative than the calculator outputs.  We ran all of our analyses using the average $K-$corrections of \citet{Mann01} and found that the excess luminosity density at $z>0.1$ is reduced by $5-10\%$.  In general, we expect the calculator outputs to be more robust estimates of the true $K-$corrections, as they were derived using a much larger sample than the $28$ local galaxies used by \citet{Mann01}, and the sample was drawn from the UKIDSS and SDSS surveys, which we use for this study.

The $E(z)$ corrections we use are of the form $E(z)=Qz$, as described in Section~\ref{absmag}, and also act to slightly reduce the observed excess in luminosity density.  
To explore the possibility of underestimated evolution, we increased the $Q$ parameter arbitrarily to investigate the effects of stronger evolution.  We found we needed to increase the evolution correction to $Q=4$ to make the average luminosity density at $z>0.1$ roughly equal to that at $z=0.05$.  This would imply that the rest-frame $K-$band light from galaxies is reduced by $\sim 40-50\%$ over the last $1-2$ Gyrs.  This amount of evolution would require galaxies to form at $z\ll1$, which is ruled out by galaxy star formation histories.  Thus, we do not expect underestimates of the $E(z)$ corrections to be a possible source of the excess luminosity density at $z>0.1$.  

\subsection{The Faint-end Slope of the Luminosity Function}
At $z>0.09$ in the UKIDSS sample we are not making a robust estimate of the faint end slope ($\alpha$) of the LF.  However, our analysis of the GAMA sample allows us to extend the faint end of the LF at $z>0.09$ by $\sim 0.7$ mags, and we find $\alpha =-1.02$ continues to be a good representation of the data out to a point where $\sim 75\%$ of the total luminosity density has been resolved.

We examined the possibility that a less negative slope than that measured at $z<0.07$ (i.e., $\alpha > -1.02$) at $z>0.09$ could account for the observed excess in luminosity density.  This is, essentially, another means of exploring the possible effects of unanticipated evolution in the LF as a function of redshift.     We found that a slope of $\alpha \approx -0.6$ imposed at $z>0.09$ (while leaving $\alpha=-1.02$ at $z<0.07$) was sufficient to suppress the observed excess at $z>0.09$.  

However, $\alpha$ and $M^*$ are correlated parameters, and changing $\alpha$ to $-0.6$ requires increasing $M^*$ by $\sim 0.2$ magnitudes to best fit the data.  Still, the best fit in this case gives a $\chi^2_{\rm{red}}\sim 4$, much worse than the value of $\chi^2_{\rm{red}}\sim 1.5$ for the case of $\alpha=-1.02, M^*=-22.15$.  Thus, the fact that the $z>0.09$ data prefer a particular value for $M^*$ already constrains $\alpha$ to be similar to that found for the low-redshift data. 

In addition, the vast majority of studies of the NIR LF from the literature measure $\alpha$ to be $\sim 0.9 -1.1$ for $0<z<0.3$ (e.g., \citealt{Cole01, Koch01, Feul03, Hill10,Keen12}).  Thus, $\alpha=-0.6$ at $z>0.09$ would not only represent some kind of extreme evolution in the LF but also be at odds with results from the literature.  Therefore, we do not consider this to be a reasonable candidate for the observed excess luminosity density at $z>0.09$.    

If we fix $\alpha$ and $M^*$ to the values derived using STY for the whole sample from $0.005<z<0.2$ ($\alpha=-0.87, M^*=-22.14$, shown in Figure~\ref{allklf}), the luminosity density at $z>0.09$ is reduced and the $z<0.07$ luminosity density is increased, somewhat reducing the tension between the low and high-redshift results.  However, this comes at the expense of a poor fit to the data in both redshift ranges ($\chi^2_{\rm{red}}\sim 3$), and a low-redshift luminosity density that is in tension with much larger all sky samples from \citet{Lava11} and \citet{Bran12}.  

Thus, in short, we find that some artificial manipulation of the shape parameters of the LF can act to reduce the apparent excess in luminosity density at $z>0.09$.  However, the simple requirements that the model parameters chosen provide the best possible fit to the data, and that the results be in harmony with much larger samples at low redshift, confirm that the shape parameters ($\alpha =-1.02, M^*=-22.15$) derived using the iterative method described in Section~\ref{kbandlf} are the most appropriate for this study.

\subsection{Spectroscopic Selection Bias}

The spectroscopy for this study comes mainly from the SDSS and is supplemented by other published redshift surveys.  Targets for SDSS spectroscopy were chosen using an $R-$band selection, and, in general, the targets for other redshift surveys from the literature were also selected optically.  Thus, there could exist a bias in the spectroscopic redshift distribution for a $K-$band selected sample for which spectroscopic targets have been selected in optical bands.  At $z\sim 0.05$, the typical $R-K$ color of galaxies is $\sim 2$, while at $z\sim 0.15, R-K \sim 1.25$.   Thus, there will be a bias against faint $K-$selected sources at low redshifts being chosen for spectroscopic follow-up in an $R-$band target selection (relative to the same selection at higher redshifts).  In principle, such a selection bias could reduce the overall $K-$band LF normalization at low redshifts.

However, the median apparent $K-$band magnitude of a galaxy at $z<0.07$ is roughly $0.75$ mags brighter than that of a galaxy at $z>0.09$ in this study, so the spectroscopic selection effect should be essentially cancelled out by the fact that nearby galaxies are intrinsically brighter on the sky, and we would expect any biases due to such  selection effects at low redshifts to be minimal.  

Still, for this reason, we restrict this study to only areas of very high spectroscopic completeness.  The minimum acceptable completeness for regions covered in this study is $90\%$, and, on average, the completeness is $\sim 95\%$ at $K_{\rm{AB}}<16.3$.   If we consider the extreme scenario in which all galaxies lacking spectroscopy are assumed to reside at $z<0.07$,  we estimate that the LF normalization at $z<0.07$ could be increased by as much as $\sim 20\%$.  However, as mentioned above, there is no physical reason that this  should be the case.  Furthermore, our measured low-redshift normalization is in very good agreement with the much larger 2M++ sample, so we believe it is accurate.  

If we assume that spectroscopic selection bias is not a problem in this study, then we can expand the area on the sky to include the entire region of overlap between the UKIDSS-LAS and the SDSS.  In doing so, the average overall completeness of the sample drops to $\sim 85\%$, but the volume is increased by roughly a factor of four (increase in area from $\sim 500$ deg$^2$ to $\sim 2000$ deg$^2$).  We performed all of the analyses described above on this full sample of $\sim 140,000$ galaxies and obtained very similar results, namely, that the $K-$band luminosity density at $z>0.1$ appears to be $\sim 1.5$ times higher than that measured locally.  

\subsection{Photometric Errors}
As noted in Section~\ref{clipping}, galaxies which subtend a large solid angle on the sky may have their Petrosian fluxes underestimated due to the upper limit of $6\arcsec$ (corresponding to a $12\arcsec$ circular aperture radius) on the size of the Petrosian aperture in the WSA pipeline.  As demonstrated in Figure~\ref{clippedvz}, this issue presents a problem at low redshifts ($z < 0.1$) where a significant fraction of objects have had their Petrosian apertures clipped, and thus their fluxes underestimated.  

In Section~\ref{clipping}, we describe a method to estimate a flux correction for galaxies where the Petrosian aperture was clipped by using a range of circular aperture magnitudes to measure a surface brightness profile for each galaxy, and then extrapolate a S\'{e}rsic profile  to the $z-$band Petrosian aperture radius given in the SDSS.  We show in Figure~\ref{counts} that making this correction brings the galaxy counts for our sample into agreement with the expected distribution taken from all-sky galaxy counts in the $K-$band.  We explored the effect of arbitrarily increasing this correction and found that a larger correction (by a factor of $\sim 2$) had the effect of overpopulating the bright end of the galaxy counts distribution.

We found that applying the aperture clipping correction factor before making the initial apparent magnitude selection in the $K-$band only increased the sample size by $\sim 0.25\%$.  Before making this aperture correction, we found that the bright end of the LF at low redshifts ($z<0.05$) appeared to fall off more steeply than expected (given the distribution seen at slightly higher redshifts).  Upon making the correction, the bright end of the LF at low redshifts was moved to brighter magnitudes and toward better agreement with higher redshift results.  

We were also able to check our aperture corrections against independent photometry from the GAMA survey, which uses the same UKIDSS-LAS imaging data as the input, and does not suffer from the same aperture clipping issue.  We found that for unclipped objects in the UKIDSS sample the rms scatter compared to GAMA Petrosian aperture photometry was $\sim 0.1$ mags and GAMA magnitudes were brighter by $\sim 0.03$ mags.  For clipped objects the scatter was the same but the systematic offset was $\sim 0.07$ mags.  We found that after the aperture clipping correction was applied the systematic offset between GAMA and UKIDSS magnitudes was the same for clipped and unclipped targets.    

We note, however, that in the lowest redshift bin presented in Figure~\ref{uklf}, the bright end of the LF still appears to fall off more steeply than at higher redshifts.  We found that arbitrarily increasing the aperture correction in this redshift bin appears to bring the bright end of the LF into better agreement with the LFs in higher redshift bins.  Thus, we conclude that we are probably underestimating the aperture clipping correction at the very lowest redshifts.  However, the aperture correction itself (even when increased arbitrarily) does almost nothing to increase the normalization of the LF.   Correcting for this effect primarily serves to shift the bright end of the LF along the abscissa.

Therefore, while our aperture clipping correction method represents a rather crude lost light correction for any individual galaxy, we believe that it provides a statistically sound means of recovering the galaxy luminosity distribution.  Furthermore, we find that the relative distribution of $K-$band luminosity density as a function of redshift presented in Figure~\ref{ukld} is not significantly altered by making this correction (or doubling it for that matter).  
    
 \subsection{Cosmic Variance}
Systematic bias due to cosmic variance is a possible source of error in any measurement made over a volume that is insufficient to average over large-scale structure in the universe.  As noted in Section~\ref{intro}, the largest structures observed in the local universe appear to be of order the size of the largest volumes surveyed to date.  Thus, it remains unclear what the upper limit on the size of structure is and what volume constitutes a representative sample of the universe.  

To address the issue of cosmic variance, we divided our sample into three subsamples denoted by the different regions on the sky shown in Figure~\ref{cover}.  As shown in Figure~\ref{ukld}, our measured luminosity density in some redshift bins varied dramatically among the three subsamples, but, in general, the overall trend toward a rising luminosity density with increasing redshift is present in all three regions on the sky.   
      
In the highest redshift bins the results from the different subsamples appear to be converging, suggesting that some kind of `scale of homogeneity' has been reached, but the substantial increase in luminosity density compared to the lowest redshift bins implies that structure may exist on scales larger than the survey volume.  We estimate the systematics due to cosmic variance in each of our six redshift bins given the rms scatter between measurements from the three subregions.  We present this error estimate in Table~\ref{table1} and graphically in Figure~\ref{ukld}b.

In the lowest redshift bin ($0.005 < z < 0.04$), our measurements of the luminosity density all agree to within the statistical $1\sigma$ errors.  In addition, we divided the 2M++ catalog up into three independent subsamples.  For simplicity, we chose the three regions to be those they identify in their study as the SDSS region ($\sim 7500$ deg$^2$), the 6DFGRS region ($\sim 17,000$ deg$^2$), and the 2MR region ($\sim 12,500$ deg$^2$).  Our measurements of the 2M++ luminosity density in these three regions (shown in Figure~\ref{ukld}a) all agree very well with one another, and with our measurement from the UKIDSS data.  In addition, these measurements agree well with independent analysis of the 2M++ catalog by \citet{Lava11} and that of a similar sample by \citet{Bran12}.  This broad agreement suggests our measurement of the local luminosity density is robust.  Interestingly, this also satisfies the requirement that any large local under-density would need to be roughly uniform about the observer (i.e., the observer must be located near the center) to avoid inducing a large CMB dipole.  

\subsection{Comparison with Mock Catalogs}
As a final check that our methods are sound, and to investigate further the expected effects of cosmic variance, we performed all the same analyses described above to estimate the relative variation in $K-$band luminosity density as a function of redshift using mock catalogs from the Millennium database online\footnote{http://gavo.mpa-garching.mpg.de/MyMillennium/}.  The Millennium database contains a variety of mock catalogs based on the Millennium \citep{Spri05} and Millennium-II \citep{Boyl09} simulations.   We used the ``Blaizot2006" all-sky catalogs, which were generated using the MoMaF code \citep{Blaz05}, as well as the ``Henriques2012" all-sky catalogs created using the virtual observing algorithm of \citet{Henr12} and the semi-analytical model of \citet{Guo11}.  

From both catalogs, we retrieved all-sky $K-$selected catalogs of galaxies at $K_{\rm{AB}}<16.3$.  We divided each all-sky catalog into 16 different regions to be considered independently.  While each of these regions of $\sim 2500$ deg$^2$ is much larger than our observed sample, we expect the effects of cosmic variance to be similar, because our observed sample spans many widely separated fields on the sky.  While it is difficult to quantify exactly how cosmic variance should scale in such a comparison, \citet{Driv10} show that for a sample made up of $N$ independent fields, cosmic variance should be reduced by a factor of $1/\sqrt{N}$.  

\begin{figure}
\begin{center}
\includegraphics[width=80mm]{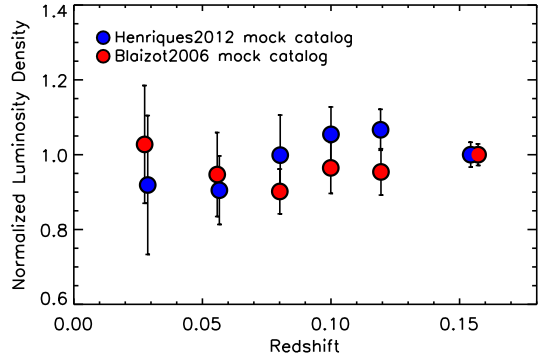}
\caption{\label{mock}  Normalized luminosity density in the $K-$band derived from two different mock catalogs using all the same methods as were used on the observed data.  In each case, we divided the all sky catalog for the $K_{\rm{AB}}<16.3$ sample into 16 sight lines, for which we calculated the luminosity density in each of the six redshift bins  individually.  The data points show the average result in each redshift bin for the Blaizot2006 catalogs (red circles) and the Henriques2012 catalogs (blue circles).  The error bars show the standard deviation of the results over the 16 sight lines in each redshift bin for both catalogs.  The results are shown normalized to the average in the highest redshift bin.  }
\end{center}
\end{figure}

We then computed the $K-$band luminosity density as a function of redshift for each of the 16 regions in both mock catalogs using the same methods as were used with the observed data, namely, applying the same $K(z)$ and $E(z)$ corrections and estimating the LF in the same way.  In Figure~\ref{mock}, we show the results of this exercise, where blue circles represent the Henriques2012 mock catalog, and red circles represent the Blaizot2006 catalogs.  The data points show the average result in each redshift bin, and all results are normalized to the average in the highest redshift bin.  The error bars show the standard deviation of the distribution of measured luminosity densities in each redshift bin for the 16 regions considered in each catalog.  

Thus, we find that a typical deviation of $\sim 10-20\%$ could be expected in this measurement between the lowest and highest redshift bins.  While this result could be further quantified by including additional or independent realizations of mock catalogs, our interest here is simply to show that when we apply our methods to simulated catalogs, we recover the expected result that the $K-$band luminosity density is not changing dramatically as a function of redshift.    

\subsection{The Hubble ``Constant"}
In all the analyses presented here, we have assumed $H_0 = 70~\rm{km~s^{-1}~Mpc^{-1}}$~(or $h_{70}=1$).   The luminosity densities presented in Figure~\ref{ukld} are directly proportional to $h_{70}$, so, for example, if the local expansion of the universe is indeed faster than the global expansion by some factor, then the apparent excess in luminosity density would be reduced by the same factor.  However, in general, this scenario would also imply less dark energy and a higher matter density, such that all these analyses would have to be repeated for the appropriate cosmology.  Here we have not attempted to quantify the effects of varying cosmological parameters, but all quantities presented are given in terms of $h_{70}$ so that the reader may readily see the impact of a variable Hubble constant. 

\section{Summary}

\label{summary}

We have presented a study of the $K-$band luminosity density at $z<0.2$.  This study is based on a  $K-$band selection of galaxies taken from the UKIDSS-LAS, where spectroscopy from the SDSS and other redshift surveys provides for a sample covering $\sim 600$ deg$^2$ on the sky that is $\sim 95\%$ spectroscopically complete at $K_{\rm{AB}}<16.3$.  The primary motivation for this study is to explore the relatively low-redshift stellar mass density distribution of the universe to test for the possibility that the nearby universe is under-dense on a scale and amplitude sufficient to introduce significant biases into locally measured cosmological observables.   This work represents the first time that the NIR luminosity density has been measured in the nearby universe with consistent methods and photometry over a wide range in redshift.  

We find that we can achieve good fits to the $K-$band galaxy LF both at low redshifts and at higher redshifts using the same LF shape parameters, $M^*$ and $\alpha$.  We find values for these LF shape parameters of $M^*=-22.15\pm 0.04$ and $\alpha=-1.02\pm 0.03$ by using the STY method and iteratively fitting $\alpha$ on the low-redshift data while holding $M^*$ fixed, then fitting $M^*$ on the higher-redshift data while  holding $\alpha$ fixed until convergence is achieved.  The result of this exercise is that we find the Schechter function normalization ($\phi^*$) at $z>0.09$ to be $\sim 1.4-1.5$ times higher than that at $z<0.07$.  

Next, we consider six redshift bins over the range $0.005<z<0.2$ to explore the $K-$band luminosity density as a function of distance in detail.  To do this, we fix $M^*$ and $\alpha$ and only fit the normalization, $\phi^*$, in each bin.  We find excellent agreement with all previous measurements of the $K-$band LF from the literature.  We confirm the tentative detection, presented in \citet{Keen12}, of a rising NIR luminosity density over the range $0.05 < z < 0.1$ and a luminosity density at $z> 0.1$ that is $\sim 1.5$ times higher than that measured locally.  

We also divide our sample into three subsamples to consider the issue of cosmic variance.  We find that, while individual measurements of the luminosity density in a given redshift bin may differ by a factor of two between subregions, the overall trend with redshift is consistent between the subregions.   At the highest redshifts considered in this sample,  the measurement in the three subsamples appears to be converging on the result that the luminosity density at $z > 0.1$ appears to be $\sim 1.5$ times higher than that measured locally.  In particular, it is worth noting that in subregions 1 and 2, which represent opposite directions on the sky, the measurements agree very well on the luminosity density in both the lowest and highest redshift bins.  

We also analyze the 2M++ catalog of \citet{Lava11} to check our low-redshift results and find good agreement with this all-sky 2MASS $K_s-$selected sample.  We divide the 2M++ catalog into three regions and find the luminosity density at $z<0.05$ to be quite uniform in different directions on the sky.

This relative agreement between widely separated sight lines, combined with the fact that the $K-$band galaxy counts from our study agree well with previously published all-sky galaxy counts, suggests that the phenomenon of a rising luminosity density with increasing distance from us may be a general trend in the nearby universe.   

We use the recently released GAMA-DR2 survey data to extend our measurement of the faint end of the LF in the higher-redshift end of the UKIDSS sample.  We find that the LF derived from the GAMA sample is in good agreement with that from the UKIDSS sample, and we use the combination of the two to resolve $\sim 75\%$ of the total luminosity density at $z>0.09$.

As NIR luminosity is a good tracer of stellar mass, and stellar mass density should trace the underlying dark matter distribution, we also compare our results with the radial density profiles of \citet{Bole11a} and \citet{Alex09}, which they claim could act to produce the apparent acceleration of the expansion of the universe observed via type Ia supernovae.  While these models are simplistic, they give a rough idea of the minimal scale and amplitude required for a local under-density to introduce significant biases in locally measured cosmological observables, such as the expansion rate.  We also compare to the analysis of \citet{Marr13}, who explore the size of a local under-density required  to explain the apparent tension between direct measurements of the Hubble constant and those recently published by the Planck collaboration.  

We conclude that local measurements of the NIR luminosity density are consistent with models that invoke a large local under-density to explain either the apparent acceleration observed via type Ia supernovae, or to explain the discrepancy between local measurements of $H_0$ and those inferred from the CMB.  A simple requirement of all these types of models is that the observer must be located somewhere near the center of the under-density to avoid seeing a large CMB dipole (some authors quote within $\sim10-15\%$ of the scale radius, e.g., \citealt{Alex09}).  While our analyses do not rigorously constrain our position with respect to local structure, we can say that the minimum requirement is met that the luminosity density appears uniformly low in all directions at $z<0.05$ and uniformly higher at $z>0.1$.  Ongoing and future redshift surveys will help to further constrain our position with respect to local large-scale structure.

\acknowledgements{We thank the anonymous referee for comments and suggestions that helped to improve this manuscript.  

We gratefully acknowledge support from the University of Wisconsin Research Committee with funds granted by the Wisconsin Alumni Research Foundation and the David and Lucile Packard Foundation (A.~J.~B.) and NSF grant AST-0709356 (L.~L.~C.).

This work is based in part on data products from the UKIRT Infrared Deep Sky Survey (UKIDSS).

This work is based in part on data products from the Galaxy And Mass Assembly (GAMA) Survey.

This publication makes use of data products from the Sloan Digital Sky Survey (SDSS). Funding for the SDSS and SDSS-II has been provided by the Alfred P. Sloan Foundation, the Participating Institutions, the National Science Foundation, the U.S. Department of Energy, the National Aeronautics and Space Administration, the Japanese Monbukagakusho, the Max Planck Society, and the Higher Education Funding Council for England. The SDSS Web Site is http://www.sdss.org/. The SDSS is managed by the Astrophysical Research Consortium for the Participating Institutions. The Participating Institutions are the American Museum of Natural History, Astrophysical Institute Potsdam, University of Basel, University of Cambridge, Case Western Reserve University, University of Chicago, Drexel University, Fermilab, the Institute for Advanced Study, the Japan Participation Group, Johns Hopkins University, the Joint Institute for Nuclear Astrophysics, the Kavli Institute for Particle Astrophysics and Cosmology, the Korean Scientist Group, the Chinese Academy of Sciences (LAMOST), Los Alamos National Laboratory, the Max-Planck-Institute for Astronomy (MPIA), the Max-Planck-Institute for Astrophysics (MPA), New Mexico State University, Ohio State University, University of Pittsburgh, University of Portsmouth, Princeton University, the United States Naval Observatory, and the University of Washington. 

This research made use of the ``K-corrections calculator'' service available at http://kcor.sai.msu.ru/.

This work has made use of NASA's Astrophysics Data System.}


\end{document}